\documentstyle[aps,preprint]{revtex}
\begin{document}
\draft
\title{Comment on "Quantum Scattering of Heavy Particles from a 10 K
Cu(111) Surface"}
\author{A. \v{S}iber and B. Gumhalter}
\address{Institute of Physics of the University, P.O. Box 304, 10001
Zagreb, Croatia}
\maketitle


In a recent Letter Althoff et al. \cite{Althoff} reported a study of
scattering of thermal Ne, Ar and Kr atoms from Cu(111)  surface in which they assessed the corresponding Debye-Waller factor
(DWF) as a function of the particle mass $m$ in a wide range of substrate temperature
$T$. The experiments were interpreted by the semiclassical
DWF theory \cite{Levi,Brenig,BK} in which 
the projectile is treated as moving on the classical
recoilless trajectory ${\bf r}(t)$ and the surface vibrations are
quantized. This gives the DWF in the form 
$I_{00}=\exp[-2W(m,T)]$ where the Debye-Waller exponent $2W$ (DWE) in the essentially one-dimensional (1D) approach of Refs. \cite{Levi,Brenig,BK} and the limit  
$T\rightarrow 0$ depends on the particle
incoming energy $E_{i}$, surface Debye temperature $\Theta_{D}$ and the
static atom-surface potential $V(z)$, but not on $m$. On the other hand, 
in the scattering regime $T\gg\Theta_{\tau}=\hbar/(k_{B}\tau)$ and  $\Theta_{\tau}<\Theta_{D}$,
where $\tau$ is the effective collision time, it scales as $2W\propto m^{1/2}T$. 
However, the experiments described in \cite{Althoff} were carried out
in the quantum scattering regime in which, as we show below, neither of the above 
semiclassical scalings holds and the semiclassical DWE significantly deviates from the exact quantum one both in the low and high $T$-limits, irrespective of the functional form of $V(z)$. Hence, the quantum scattering data  \cite{Althoff} cannot be reliably interpreted by the semiclassical but rather by the quantum theory.

To substantiate the above statements we carry out fully
three-dimensional quantum \cite{HAS,Siber,Braun}
and semiclassical \cite{BK} calculation of the DWF relevant to the experiments of Ref.
\cite{Althoff}. We start from the quantum DWE (c.f. Ref. 
\cite{Siber}, Eq. (3)) in which we also include prompt sticking
processes because of their large contribution to the quantum DWF and use theoretical $V(z)$'s from \cite{Althoff} which produce closer fit of the semiclassical DWF to the data. The quantum DWF has the correct semiclassical limit \cite{BGL,HAS} which enables pinpointing the break down of the semiclassical
description.  
A comparison of the measured \cite{Althoff} and calculated DWF's 
shown in Fig. \ref{fig1} reveals general agreement between  the measured and quantum and not the semiclassical values, and thereby establishes the validity of the quantum approach. 
A systematic
small underestimate at very low $T$ appears because our results are uncorrected for $T$-independent diffuse elastic scattering. 
The breakdown of the semiclassical approach is illustrated in 
Fig. \ref{fig2} which shows the $m$-dependence of the quantum  and semiclassical DWE's calculated in the low (main panel) and high (inset) $T$-limit for fixed $V(z)$ and $E_{i}$. 
It is seen that the semiclassical  1D scaling results for the DWE
are reached for masses which largely exceed those of the scattered atoms (classical limit). Hence, although our calculations support the
conjectures of Ref. \cite{Althoff} on the $m$-dependence
of the DWF for heavy particle-surface scattering in the {\em classical}
limit, they also demonstrate
that the semiclassical approach invoked to interpret the data \cite{Althoff} is not yet applicable in the studied scattering regime and that the {\em quantum} approach should be used instead.

\vskip 3.8 cm

\begin{figure}
\caption{Calculated temperature dependence of the quantum (full curves) and semiclassical (dashed curves) Debye-Waller factor
for Ne, Ar and Kr scattering from Cu(111) and
experimental results \protect\cite{Althoff} (full symbols). } 
\label{fig1}
\end{figure}

\vskip 4.5 cm

\begin{figure}
\caption{Dependence of quantum 
(full curves) and semiclassical Debye-Waler exponent (dashed curves) on 
particle mass for fixed $V(z)$ and $E_{i}$ in the low and high temperature limit. }
\label{fig2}
\end{figure} 

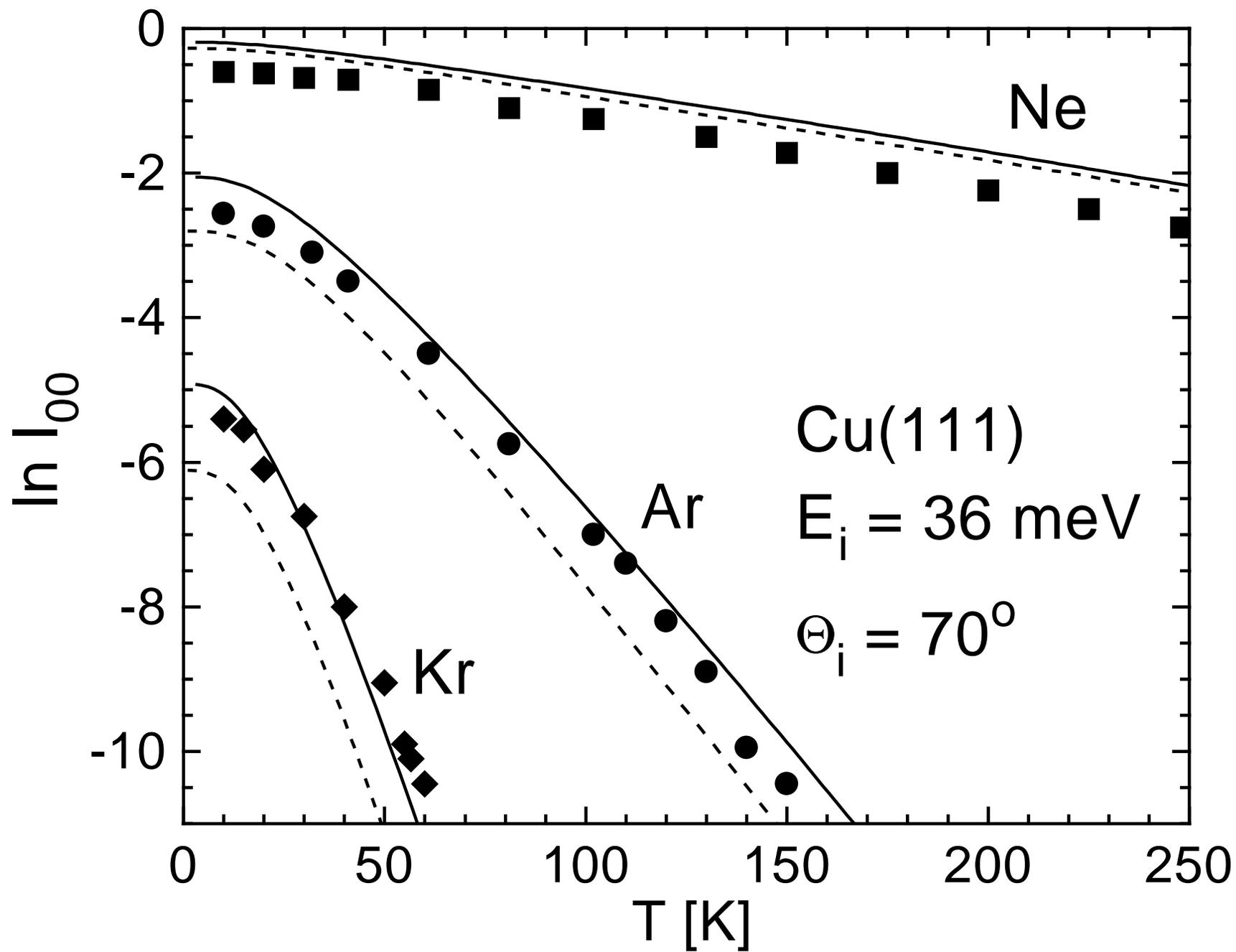
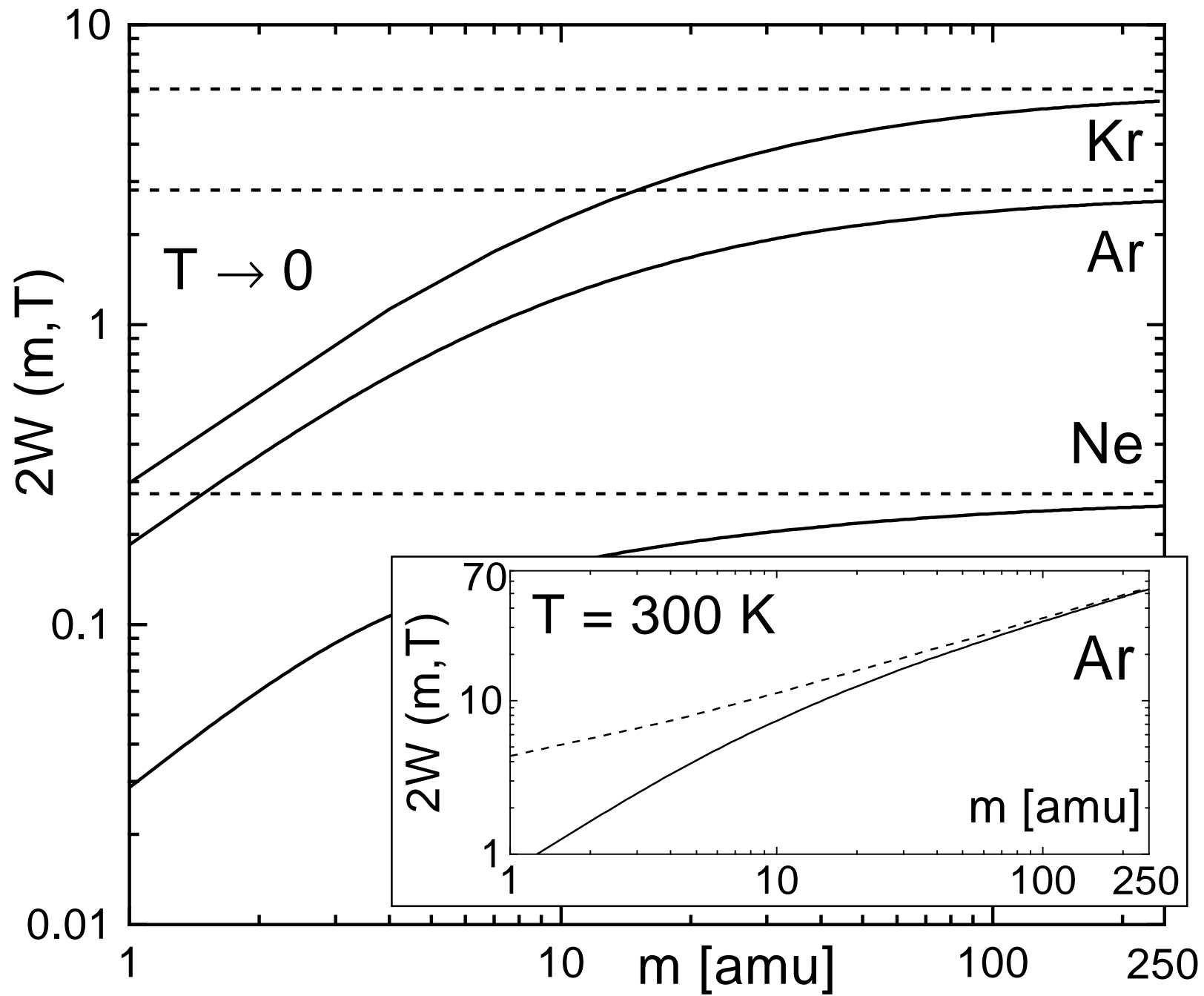
\end{document}